# Efficient GPU-computing simulation platform JAX-PF for differentiable phase field model

Fanglei Hu[a], Jiachen Guo[b,c], Stephen Niezgoda[d], Wing Kam Liu[b,c], Jian Cao[a*]

[a] Department of Mechanical Engineering, Northwestern University, Evanston, IL 60208, USA

[b] Theoretical and Applied Mechanics, Northwestern University, Evanston, IL, 60208, USA

[c] HIDENN-AI, LLC, 1801 Maple Ave, Evanston, 60201, IL, USA

[d] Department of Materials Science and Engineering, The Ohio State University, Columbus, OH 43210, USA

* Corresponding author; Email address: jcao@northwestern.edu (Jian Cao).

**Abstract**

We present JAX-PF, an open-source, GPU-accelerated, and differentiable Phase Field (PF) software package, supporting both explicit and implicit time stepping schemes. Leveraging the modern computing architecture JAX, JAX-PF achieves high performance through array programming and GPU acceleration, delivering ~5× speedup over PRISMS-PF with MPI (24 CPU cores) for systems with ~4.19 million degrees of freedom using explicit schemes, and scaling efficiently with implicit schemes for large-size problems. Furthermore, a key feature of JAX-PF is automatic differentiation (AD), eliminating manual derivations of free-energy functionals and Jacobians. Beyond forward simulations, JAX-PF demonstrates its potential in inverse design by providing sensitivities for gradient-based optimization. We demonstrate, for the first time, the calibration of PF material parameters using AD-based sensitivities, highlighting its capability for high-dimensional inverse problems. By combining efficiency, flexibility, and full differentiability, JAX-PF offers a fast, practical, and integrated tool for forward simulation and inverse design, advancing co-designing of material and manufacturing processes and supporting the goals of the Materials Genome Initiative.

1. INTRODUCION

Advanced manufacturing has emerged as a transformative paradigm in modern industry, integrating advanced digital technologies with traditional processing methods to co-designing of material and manufacturing processes [1-4]. A key component of this paradigm is the Integrated Computational Materials Engineering (ICME), which combines tools across multiple scales—including molecular dynamics simulations, crystal plasticity, and continuum-scale constitutive modeling—to systematically link processing, microstructural evolution, and resulting material properties [5-7]. This approach has significantly advanced the forward design of materials and processes. Within ICME, the Phase Field (PF) method has become a key modeling tool for predicting the evolution of microstructures and their properties at nano- to mesoscopic scales [8]. Based on continuum mass density functional theory, PF method introduces different PF variables to distinguish different phases by their local characteristics (e.g., atomic order, concentration, lattice strain) and eliminates the need for explicit interface tracking [9]. These features make PF particularly well-suited for simulating complex phase transformations in metallic materials, including dendrite solidification [10, 11], spinodal decomposition [12], recrystallization [13, 14], diffusionless martensite transformation [15-17], precipitation [18-22], crack evaluation [23-25], and so on. By capturing these mechanisms, PF modeling provides valuable insights for designing high-performance materials and guiding manufacturing processes. A detailed review of PF models and their applications can be found in the literature [8, 26-29].

For co-designing of material and manufacturing processes, inverse approaches that directly derive microstructures and processing parameters from desired properties are particularly attractive [30-32]. Inverse design treats target properties as inputs to determine the initial geometry, microstructure, and subsequent processing conditions [33]. Existing optimization strategies generally fall into two categories: gradient-free and gradient-based. Gradient-free methods, such as genetic algorithm [34] and Bayesian optimization [35], explore the design space through random sampling and heuristic searches. While gradient-free methods do not require gradient information, they often converge slowly and may suffer from significant computational cost and low accuracy in high-dimensional inverse problems. In contrast, gradient-based optimization exploits gradient information to guide parameter updates, enabling faster convergence and making it more suitable for large and complex design spaces. For example, thermomechanical processing in forging with targeted final microstructure and properties requires optimization over high-dimensional parameters—such as thermal history, heat treatment, and loading path at each material point—making gradient-based methods indispensable [33, 36]. Despite this potential, the application of gradient-based optimization to PF inverse design remains largely unexplored due to four key challenges. First, PF models often involve multiple PF variables, case-specific free energy functionals, and coupled internal/external fields, making their integration into numerical frameworks a lengthy and complex process. Second, sensitivities—the gradients of objective functions with respect to design parameters—are generally unavailable in conventional PF codes. Third, inverse design requires repeated forward simulations in iterations, but PF simulations have mismatch of length scales in nature and computationally intensive, often consuming millions of core hours on High performance computing (HPC) platforms. Finally, while PF methods can quantitatively link processing and microstructure, effective integration with other computational tools, such as crystal plasticity and continuum plasticity models, is still needed to fully connect microstructure to mechanical properties. A tightly coupled ecosystem for multi-scale simulations is therefore needed.

To address these challenges, we developed `JAX-PF`, built upon our recent open-source finite element method library, `JAX-FEM` [37]. `JAX-PF` provides an efficient, flexible, ease-of-use, and collaborative platform for general-purpose differential PF simulation tools, which is suitable for inverse design. The following sections will review the current PF software and provide the background and motivation for our developed software.

First, PF models are almost exclusively solved numerically, most commonly using finite difference method (FDM), finite volume method (FVM), fast Fourier transform (FFT), or finite element methods (FEM), as detailed in Section "Methods". While we also provide finite difference implementations for completeness, this work primarily focuses on FEM. FEM is well-suited for predicting microstructural evolution in complex manufacturing processes [38, 39], but its implementation faces two major challenges: (i) Implicit FEM schemes for PF typically adopt predictor–corrector updates for state variables, including multi-PF and multi-physics variables (e.g., displacement and temperature fields and so on), solved iteratively with the Newton–Raphson method until convergence. This method requires the algorithmic tangent modulus for calculating the Jacobian matrix. For high-fidelity PF models involving coupled partial differential equations (PDEs) for multiple PF parameters and continuous fields (e.g., temperature, strain, composition), derivation of case-by-case Jacobian matrix in Newton–Raphson iterations becomes prohibitively complex; and (ii) PF models inherently rely on free energy functionals of one or more density functions or PF variables to describe the evolution of mesoscale microstructure. Derivatives of these energy terms are essential for the solution but are often cumbersome to compute, particularly in advanced models such as the Kim–Kim–Suzuki (KKS) formulation [40]. *JAX-PF addresses these difficulties* through automatic differentiation (AD) within `JAX-FEM` [37], enabling machine-precision computation of derivatives of energy terms and Jacobian matrices without manual derivations [41]. Although the use of AD for nonlinear mechanics, crystal plasticity, and phonon Boltzmann Transport Equation has been explored in the last few years [42-45], its application in PF modeling remains largely unexplored. Again, the use of AD in `JAX-PF` shifts the burden of computing the derivatives and facilitates the realization of multi-physics and multi-PF variables modeling for forward analysis.

Second, accurate and efficient sensitivity analysis—i.e., the gradient of objective functions with respect to design parameters—is essential for enabling gradient-based optimization in inverse problems [46, 47]. For example, to quantify the influence of material parameters such as elastic modulus, diffusion coefficient, or kinetic mobility, sensitivity analysis provides the derivatives of microstructural features with respect to these inputs. However, in PF models, the strong nonlinearity arising from the coupling of multiple PF variables, complex free energy functionals, and internal/external fields makes such derivations highly complex. It requires considerable effort, as the composition–processing–microstructure–property relationships are implicitly defined. Although `JAX-FEM` provides automatic sensitivity analysis functions, it faces two main challenges when applied to PF: (i) Differentiating through thousands or even millions of explicit time steps using reverse-mode AD is prohibitively expensive, since all intermediate states must either be stored or recomputed during the backward pass; and (ii) The existing `JAX-FEM` implementation supports only relatively simple nonlinear models, such as J2 plasticity model [48], while PF systems involve complex couplings among different PF variables and external fields. *To overcome these limitations*, `JAX-PF` introduces an implicit time-stepping scheme, which allows much larger time steps and dramatically shortens the trajectory length for AD. Additionally, customized differentiation rules based on the adjoint method [49] are incorporated in `JAX-PF` to compute sensitivities efficiently in complex nonlinear PF models. Compared with finite-difference numerical derivatives, AD in `JAX-PF` delivers machine-precision accuracy while maintaining computational efficiency, particularly in high-dimensional parameter design spaces.

Third, high-fidelity PF simulations are computationally expensive. They require solving a system of coupled PDEs for multiple field variables on sufficiently fine spatial discretization to resolve microstructural features such as phase interfaces. When PF is combined with inverse design for processing/microstructure design, repeated forward simulations in iterations further enhance the computational burden. As a result, computational performance and scalability are critical concerns. Some existing open-source PF software [50], such as `MOOSE` [51], `PRISMS-PF` [52], and `Open-Phase` [53], employ parallel CPU implementations that scale with increasing processor counts. However, even with 32 to 64 CPU cores, large-scale simulations often require days of wall time, and access to supercomputers with hundreds of processors is not always practical for most researchers. *To overcome these limitations*, general-purpose graphical processing units (GPUs) provide a compelling alternative, as they are specialized for highly parallel, compute-intensive tasks and offer significantly greater throughput than CPUs [54]. To the best of knowledge of the authors, `MicroSim` [55] is one of the few open-source PF solvers supporting both CPU and GPU execution, but it is limited to FDM, FVM, and FFT methods. In contrast, GPU-accelerated PF solver based on FEM remains largely unexplored, despite FEM being particularly powerful for handling complex geometries, heterogeneous material properties, nonlinear constitutive laws, and multi-physics coupling. `JAX-PF` fills this gap by leveraging the `XLA` (Accelerated Linear Algebra) backend of `JAX` [41], delivering highly competitive GPU performance with minimal user effort. Furthermore, `JAX-PF` implements vectorized operations across multiple PF variables (e.g., order parameters in grain growth) based on `jax.vmap`, a core function in `JAX`, which further exploits GPU parallelism. Together, these features enable `JAX-PF` to substantially accelerate PF simulations, thereby expanding their applicability to broader materials design and manufacturing problems.

Finally, the growing interest in process–structure–property relationships—which critically depend on mesoscale material characteristics and are crucial for both forward and inverse analyses—has created a strong demand for integrated platforms with unified solutions. A promising direction lies in multiscale simulations that couple PF models, describing process–microstructure relations, with crystal plasticity finite element methods (CPFEM), capturing microstructure–property relations. The PRISMS Center, for example, has made large progress in this area by linking `PRISMS-PF` [56] with `PRISMS-Plasticity` [57]. However, this workflow relies on separate software packages and requires transferring data back and forth, which can be cumbersome and computationally inefficient. In contrast, `JAX-PF` is built on the same `JAX-FEM` [37] foundation as `JAX-CPFEM` [58], our recently developed GPU-accelerated differentiable CPFEM framework. This shared infrastructure enables seamless integration of PF and CPFEM within a single ecosystem, allowing users to perform fully coupled multiscale simulations—such as dynamic recrystallization [59, 60], crack propagation [61, 62], and martensite transformation [63]—with high efficiency while maintaining differentiability for gradient-based optimization and inverse design. Also, built on `JAX`, coupling of `JAX-PF` and `JAX-CPFEM` is straightforward with a simple Python interface and access to all its machine learning functionalities, offering user-friendly experience for users and developers, enabling the efficient handling of large-scale complex problems.

In summary, `JAX-PF` is built on the solid foundation of conventional PF, where microstructure evolution is described using two classes of field variables [9]: non-conserved structural order parameters governed by the Allen–Cahn equation, and conserved density variables governed by the Cahn–Hilliard equation, as detailed in the section "Methods" and the comprehensive reviews of PF [8, 26-29]. To validate and demonstrate the platform, four benchmark problems are provided, including Allen–Cahn, Cahn–Hilliard, coupled Allen–Cahn and Cahn–Hilliard, and Eshelby inclusion for lattice misfit in solid-state phase transformations [64], each implemented with both explicit and implicit time integration. On top of these benchmarks, five representative applications are included, ranging from solidification to solid-state transformations: grain growth, static recrystallization, spinodal decomposition, precipitation with the Wheeler–Boettinger–McFadden (WBM) model [65], and precipitation with the Kim–Kim–Suzuki (KKS) model [40]. We want to highlight the following four features that differentiate `JAX-PF` differs from other PF software:

1. **Ease-of-use**: Leveraging capability of automatic differentiation (AD) in `JAX`, `JAX-PF` automatically generates Jacobians matrix for Newton–Raphson method and derivatives of different energy terms with machine precision, enabling the realization of multi-physics and multi-variable PF models.

2. **Automatic Sensitivity**: Implicit time integration with customized adjoint-based AD enables efficient gradient-based optimization and inverse design of strongly nonlinear PF systems.

3. **High-performance GPU-acceleration**: Through the XLA backend and vectorized operations, `JAX-PF` delivers competitive GPU performance, drastically reducing computational time compared to CPU-based solvers.

4. **Unified multiscale ecosystem with `JAX-CPFEM`**: Built on the same `JAX-FEM` foundation, `JAX-PF` integrates seamlessly with `JAX-CPFEM` to enable coupled process–structure–property simulations (e.g., dynamic recrystallization), while preserving full differentiability for optimization and design.

The remainder of the paper is organized as follows. Section 2 reviews the governing PF formulations, free-energy models, and numerical implementation. It also outlines key features of `JAX-PF` that distinguish it from conventional PF implementations, including automatic constitutive modeling, array programming, and built-in sensitivity analysis. Section 3 presents four benchmark PF problems: the Allen–Cahn equation for non-conserved variables, the Cahn–Hilliard equation for conserved variables, the coupled Allen–Cahn and Cahn–Hilliard equations, and the Eshelby inclusion problem for lattice misfit in solid-state transformations. The computational performance of the explicit solver of `JAX-PF` is compared with `PRISMS-PF`, a widely used open-source PF software. At the same time, the scaling of the implicit solver is also demonstrated. This section further introduces sensitivity analysis using differentiable programming and compares the accuracy of AD with finite-difference numerical derivatives across all benchmarks. Building on these verified sensitivities, we propose a pipeline for inverse design and demonstrate its effectiveness through the calibration of PF material parameters in Mg–Nd alloys, using AD-based sensitivities within a gradient-based optimization framework. The scheme of notation and other technical information is compiled in Supplementary Note 1.

## 2. METHODS

This section reviews the formulations of Phase Field (PF) method, including the benchmarks, free-energy formulation, and numerical implementation. Key innovations of `JAX-PF` over conventional PF include automatic differentiation, array programming, and automated sensitivity evaluation.

### 2.1 Fundamental Theory of Phase-field Model

Extensive reviews cover both the foundations of the PF methodology and its application areas [8, 26-29]. The PF method provides a general framework for modeling microstructure evolution, with governing equations that can be broadly categorized into two types, as detailed by Fried and Gurtin [66, 67]. The Allen–Cahn equation describes the evolution of non-conserved order parameters $\eta$, capturing processes such as phase transitions driven by bulk free energy reduction and domain coarsening driven by interfacial energy minimization, as shown in Eq. (1). The Cahn–Hilliard equation governs conserved field variables, such as concentration or atomic density, where mass conservation is required (Eq. (2)). In addition, many solid-state phase transformations, including martensitic transformation [68, 69], rafting [70, 71], and precipitation [72, 73], involve elastic interactions caused by lattice misfit between matrix and secondary phases. This motivates the inclusion of an Eshelby inclusion benchmark, which evaluates the displacement field around a misfitting secondary phase (Eqs. 3–4).

$$\frac{\partial \eta_i}{\partial t} = -M_\eta \frac{\delta F}{\delta \eta_i}, \tag{1}$$

$$\frac{\partial c_i}{\partial t} = -\nabla \cdot \left(-M_c \frac{\delta F}{\delta c_i}\right), \tag{2}$$

$$\nabla \cdot \boldsymbol{\sigma} = 0, \tag{3}$$

$$\boldsymbol{\sigma} = \boldsymbol{C}:(\boldsymbol{\varepsilon}_{tot} - \boldsymbol{\varepsilon}_0), \tag{4}$$

where $F$ is the total free energy function generally includes local free-energy function, gradient energy, and so on. $M_\eta$ denotes the structural relaxation coefficient, and $M_c$ is the atomic mobility between different components. $\frac{\delta F}{\delta \eta_i}$ and $\frac{\delta F}{\delta c_i}$ denote the thermodynamic driving force. Eq. (3) is the mechanical equilibrium condition, and $\boldsymbol{\sigma}$ is the elastic stress given by Eq. (4). $\boldsymbol{\varepsilon}_{tot}$ denotes the total strain, and $\boldsymbol{\varepsilon}_0$ is the total eigenstrain. These governing equations describe the evolution of microstructures driven by minimization of the total free energy $F$, which may include bulk chemical energy, interfacial energy, elastic strain energy, magnetic energy, electrostatic energy, and contributions from external fields such as applied stress, temperature, or magnetic fields. For the Allen-Cahn and Cahn-Hilliard benchmarks, the total free energy of an inhomogeneous microstructure is written as:

$$F_{Allen-Cahn}(\eta_i, \nabla\eta_i) = F_{chem} + F_{grad} = \int_V \left( f_{chem}(\eta_i) + \sum_{i=1}^n \frac{\kappa_\eta}{2} |\nabla\eta_i|^2 \right) dV, \quad (5)$$

$$F_{Cahn-Hilliard}(c_i, \nabla c_i) = F_{chem} + F_{grad} = \int_V \left( f_{chem}(c_i) + \sum_{i=1}^m \frac{\kappa_c}{2} |\nabla c_i|^2 \right) dV, \quad (6)$$

where $f_{chem}$ is the local chemical free energy density and $f_{grad}$ represents the interfacial energy density arising from the gradient energy terms that are nonzero only at and around the interface. Here, $\kappa_\eta$ and $\kappa_c$ are gradient coefficients, and $n$ and $m$ denote the number of non-conserved and conserved variables, respectively. For coupled Allen-Cahn and Cahn-Hilliard systems, we neglect the gradient terms for composition ($\nabla c_i$) and express the total free energy term as:

$$F_{Coupled}(c_i, \eta_i, \nabla\eta_i) = F_{chem} + F_{grad}$$
$$= \int_V \left( f_\alpha(c_i)(1 - h(\eta_i)) + f_\beta(c_i)h(\eta_i) + \sum_{i=1}^n \frac{\kappa_\eta}{2} |\nabla\eta_i|^2 \right) dV, \quad (7)$$

where $h(\eta_i)$ is a monotonic interpolation function of the structural order parameter varying from 0 to 1. Additional contributions, such as elastic strain energy, plastic deformation, or magnetic/electric fields, can be incorporated to model specific physical processes. For instance, elastic stain energy was considered in the total free energy for the precipitation in Mg-Nd alloys, given by:

$$f_{ela} = \frac{1}{2} C_{ijkl}(\varepsilon_{ij} - \varepsilon_{ij}^0)(\varepsilon_{kl} - \varepsilon_{kl}^0), \quad (8)$$

where $C_{ijkl}$ is the spatially dependent elastic modulus tensor, $\varepsilon_{ij}$ is the small strain tensor, and $\varepsilon_{ij}^0$ is the composition-dependent stress-free strain transformation tensor corresponding to each structural order parameter. See ref. [64, 74-76] for more details.

## 2.2 Numerical Aspect of PF models

The Cahn–Hilliard equation involves fourth-order spatial derivatives. When cast in weak form, these fourth-order derivatives introduce second-order derivatives of the trial and test functions, which cannot be directly handled using standard Lagrange finite element bases. To address this, we reformulate the equation as a system of two coupled second-order PDEs:

$$\frac{\partial c_i}{\partial t} - \nabla \cdot (M_c \nabla \mu) = 0 \quad \text{in } \Omega, \tag{9}$$

$$\mu - \frac{df_{chem}}{dc} + \frac{\kappa_c}{2} |\nabla c_i|^2 = 0 \quad \text{in } \Omega, \tag{10}$$

where the primary unknowns are the conserved variable $c$ and the chemical potential $\mu$. This mixed formulation avoids the need for higher-order elements and allows the use of standard finite element discretization. Once discretized in space, the equations can be advanced in time using either explicit or implicit schemes.

Once the governing equations are spatially discretized, such as Allen-Cahn and Cahn-Hilliard equations, we obtain a system of ordinary differential equations (ODEs) or differential–algebraic equations (DAEs) of the form:

$$\boldsymbol{M}\frac{d\boldsymbol{U}}{dt} + \boldsymbol{r}(\boldsymbol{U}, \boldsymbol{\theta}) = 0, \tag{11}$$

where $\boldsymbol{M} \in \mathbb{R}^{N \times N}$ represents the mass matrix, $\boldsymbol{U} \in \mathbb{R}^N$ represents the vector of PF solution and multi-physics variables for DOF (e.g. $\eta, c, \mu, \boldsymbol{u}$), and $\boldsymbol{r}(\boldsymbol{U}, \boldsymbol{\theta}) \in \mathbb{R}^N$ denotes the discretized weak form depending on the solution state and material parameters.

In explicit schemes, the solution at the next time step is computed directly from the current state:

$$\boldsymbol{U}^{n+1} = \boldsymbol{U}^n - \Delta t \boldsymbol{M}^{-1} \boldsymbol{r}(\boldsymbol{U}^n, \boldsymbol{\theta}). \tag{12}$$

This method is straightforward to implement and computationally inexpensive per step. However, the stability condition requires the time step $\Delta t$ to be very small, especially for stiff systems such as PF models with fine spatial resolution or strong coupling between variables. Consequently, explicit schemes often demand a large number of time steps to reach physically meaningful time scales.

In implicit schemes, the unknown state at the next time step appears inside the residual $\boldsymbol{r}$:

$$\boldsymbol{M}\frac{\boldsymbol{U}^{n+1} - \boldsymbol{U}^n}{\Delta t} + \boldsymbol{r}(\boldsymbol{U}^{n+1}, \boldsymbol{\theta}) = 0. \tag{13}$$

This method requires solving a nonlinear system at each time increment, typically using Newton–Raphson iterations. While each step is more computationally intensive, implicit schemes are unconditionally stable and allow significantly larger time steps, making them attractive for long-time simulations. Based on array programming and automatic differentiation (AD), `JAX-PF` can solve such second-order elliptic partial differential equation by Newton-Raphson method (see [37] for details). A flow chart of the solver loop for both explicit and implicit time stepping scheme is given in Figure *1*. At each time step, a single cycle of the inner loop is executed to solve the governing equations, or once in the case of a time-independent problem.

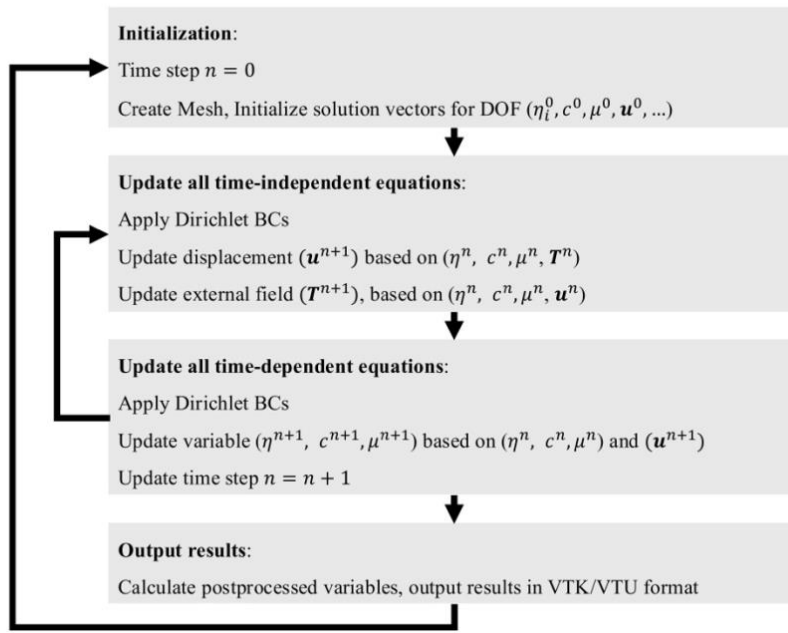

Figure 1. Flowchart of the solver loop for both explicit and implicit time stepping schemes.

## 2.3 Automatic Modeling

Following standard practice in nonlinear finite element analysis, the Newton–Raphson scheme is employed to solve the global residual equation, such as Eq. (13), with the full residual defined in Eq. (14). At each integration point, the update of the solution vector $U$ requires the Jacobian matrix $\frac{\partial R}{\partial U}$. Constructing these analytical derivatives manually is highly tedious due to the strong nonlinearity of the governing equations. In many cases, the exact Jacobian becomes intractable, particularly when: (1) multiple PF variables are involved, such as when chemical reactions and structural phase transformations are coupled at the same location; (2) multi-physics processes are applied, including heating, elasticity, or diffusion; (3) multiscale simulations are considered, where local interactions between different physical models must be captured. To address these challenges, on top of `JAX-FEM`, `JAX-PF` leverages `AD` capabilities to automatically evaluate the Jacobian matrix $\frac{\partial R}{\partial U}$, making the implementation both efficient and generalizable. The global residual takes the form as below:

$$R(U^{n+1}, \theta) = M \frac{U^{n+1} - U^n}{\Delta t} + r(U^{n+1}, \theta). \tag{14}$$

In addition to computing Jacobians for the global residual, AD in `JAX-PF` can also be directly applied to evaluate derivatives of complex free-energy functionals. This capability is particularly critical in advanced models such as the Kim–Kim–Suzuki formulation, where multiple phase concentrations are introduced as additional variables to ensure thermodynamic consistency across phases. In these models, the free-energy landscape involves nonlinear couplings between conserved and non-conserved fields, and accurate evaluation of higher-order derivatives, such as chemical potential, driving forces, and mobility terms, is indispensable. Traditionally, these derivatives must be derived and implemented manually, a process that is both tedious and error-prone. With AD, `JAX-PF` can automatically generate all required derivatives with machine precision, drastically simplifying implementation while ensuring accuracy. This advantage not only accelerates the development of new free-energy models but also makes the framework broadly accessible to researchers, allowing them to focus on physical model design rather than derivative bookkeeping.

**2.4 Array programming**

Through automatic linearization, `JAX-PF` operates directly on the weak form, with each load step linearized via AD. Rather than relying on conventional element-wise for-loops in `Fortran` or `C/C++`, `JAX-PF` adopts an array programming paradigm—akin to NumPy [24] —to achieve computational efficiency. A key feature is the use of `jax.vmap`, which vectorizes computations across all integration points and finite element cells simultaneously. This not only accelerates residual and Jacobian evaluation but also makes the implementation highly scalable on GPUs.

Beyond element-level vectorization, `JAX-PF` further extends this concept to multiple phase-field variables governed by similar evolution equations and energy functional. For example, in grain growth simulations, each grain is represented as a distinct order parameter governed by the similar Allen–Cahn formulation. `JAX-PF` updates all of grain order parameters on each quadrature point simultaneously through vectorization, fully exploiting GPU parallelism and enabling simulations with unprecedented numbers of grains. This capability substantially broadens the applicability of `JAX-PF` to large-scale microstructure evolution problems that were previously constrained by computational cost.

### 2.5 Automatic Sensitivity

Inverse design plays a central role in modern materials engineering, encompassing processing strategies, microstructure tailoring, and property optimization. Such problems can be rigorously expressed as PDE-constrained optimization (PDE-CO) formulations [77]. In this study, we follow the discretize-then-optimize approach, and the discretized PDE-CO problem of calibration is formulated as

$$\min_{\boldsymbol{U} \in \mathbb{R}^N, \boldsymbol{\theta} \in \mathbb{R}^M} O(\boldsymbol{U}, \boldsymbol{\theta}) \quad (15)$$
$$\text{s.t.} \quad \boldsymbol{C}(\boldsymbol{U}, \boldsymbol{\theta}) = 0,$$

where $\boldsymbol{U} \in \mathbb{R}^N$ represents the PF solution vector for DOF, $\boldsymbol{\theta} \in \mathbb{R}^M$ denotes the materials parameters, $\boldsymbol{C}(\cdot, \cdot) : \mathbb{R}^N \times \mathbb{R}^M \to \mathbb{R}^N$ is the constraint function representing the discretized governing PDE and should be regarded as the PF forward simulations predicting the microstructural features, and the objective function as $O(\boldsymbol{U}, \boldsymbol{\theta}) : \mathbb{R}^N \times \mathbb{R}^M \to \mathbb{R}$. By embedding the PDE constraint, the objective function can be re-formulated as

$$\hat{O}(\boldsymbol{\theta}) := O(\boldsymbol{U}(\boldsymbol{\theta}), \boldsymbol{\theta}), \quad (16)$$

where $\boldsymbol{U}(\boldsymbol{\theta})$ is the implicit function that arises from solving the PDE. Therefore, the discretized PDE-CO problem becomes

$$\boldsymbol{\theta}^* = arg\min_{\boldsymbol{\theta} \in \mathbb{R}^M} \hat{O}(\boldsymbol{\theta}). \quad (17)$$

To solve this optimization for the optimal parameter set $\boldsymbol{\theta}^*$, two strategies are commonly employed: gradient-free and gradient-based approaches. While gradient-free methods do not need to consider any information about the gradient of the objective function, they rely on random sampling and often converge slowly, especially in high-dimensional parameter spaces. On the other hand, gradient-based methods, leverage sensitivity information to guide parameter updates, offering much faster convergence and is better suited to high-dimensional problems [78]. Their main limitation is the need for accurate and efficient gradient evaluation—i.e., sensitivities of the objective function with respect to material parameters. For PF models, where nonlinear couplings among multiple variables make analytic derivations complex and error-prone, this sensitivity computation represents a central bottleneck.

To overcome these challenges, we introduce the concept of differentiable PF modeling, in which arbitrary end-to-end derivatives are made available within the proposed `JAX-PF` pipeline. This capability enables the direct application of gradient-based optimization methods to solve inverse problems such as parameter calibration. The total derivative of $\hat{O}(\boldsymbol{\theta})$ with respect to parameter variables $\boldsymbol{\theta}$ can be calculated through chain rules, see Supplementary Note 2 for details, and is formulated as

$$\boldsymbol{J} = \frac{\partial \hat{O}(\boldsymbol{\theta})}{\partial \boldsymbol{\theta}}, \tag{18}$$

where $\boldsymbol{J}$ is the gradient.

With its implicit solver, `JAX-PF` computes derivatives in a fully automated manner, providing both efficiency and accessibility for general users. Reliable sensitivity evaluation further serves as the foundation for gradient-based optimization, thereby enabling the application of such algorithms to inverse problems.

## 3. RESULTS and DISCUSSIONS

As outlined in the introduction, open-source frameworks for phase-field (PF) modeling have gained increasing attention. To evaluate the performance of `JAX-PF`, we benchmarked it against reference results reported on the `PRISMS-PF` website [56] using four examples: Allen–Cahn, Cahn–Hilliard, coupled Allen–Cahn and Cahn–Hilliard, and Eshelby inclusion for lattice misfit in solid-state transformations (Eqs. 1-4). Each case was solved with different levels of mesh resolution under an explicit time stepping scheme, using `JAX-PF` on GPU and `PRISMS-PF` on CPUs with MPI-based parallelization. To further highlight efficiency, we compared the performance of `JAX-PF` on 1 GPU with that of `PRISMS-PF` running on 24 CPU cores. The scalability of implicit time integration for these benchmarks was also examined.

In the following section, "AD-based Sensitivity Analysis", we demonstrate the differentiability of `JAX-PF` by performing and verifying sensitivity analyses for the four benchmarks using AD. We then illustrate the flexibility of `JAX-PF` by applying it to five representative problems, covering diverse physical phenomena to showcase the breadth of the framework. Finally, in Section "Inverse Problems in PF via AD-based Sensitivities", we present a case study on the calibration of PF model parameters for precipitation in Mg–Nd alloys, formulated as a multi-objective high-dimensional optimization problem enabled by AD-based sensitivities.

### 3.1 Validation of JAX-PF

To validate `JAX-PF`, we performed benchmark comparisons against `PRISMS-PF`, considering both explicit and implicit time-integration schemes in `JAX-PF` and explicit schemes in `PRISMS-PF`. The initial conditions and final solutions for representative cases are shown in Figure 2. These systems of PDEs represent simplified versions of models widely used for solidification and solid-state transformations. Figure 2 demonstrates that `JAX-PF` matches `PRISMS-PF` benchmarks, thereby confirming the accuracy of the framework. Full descriptions of the benchmark setup and numerical details are provided in the Supplementary Note 3.

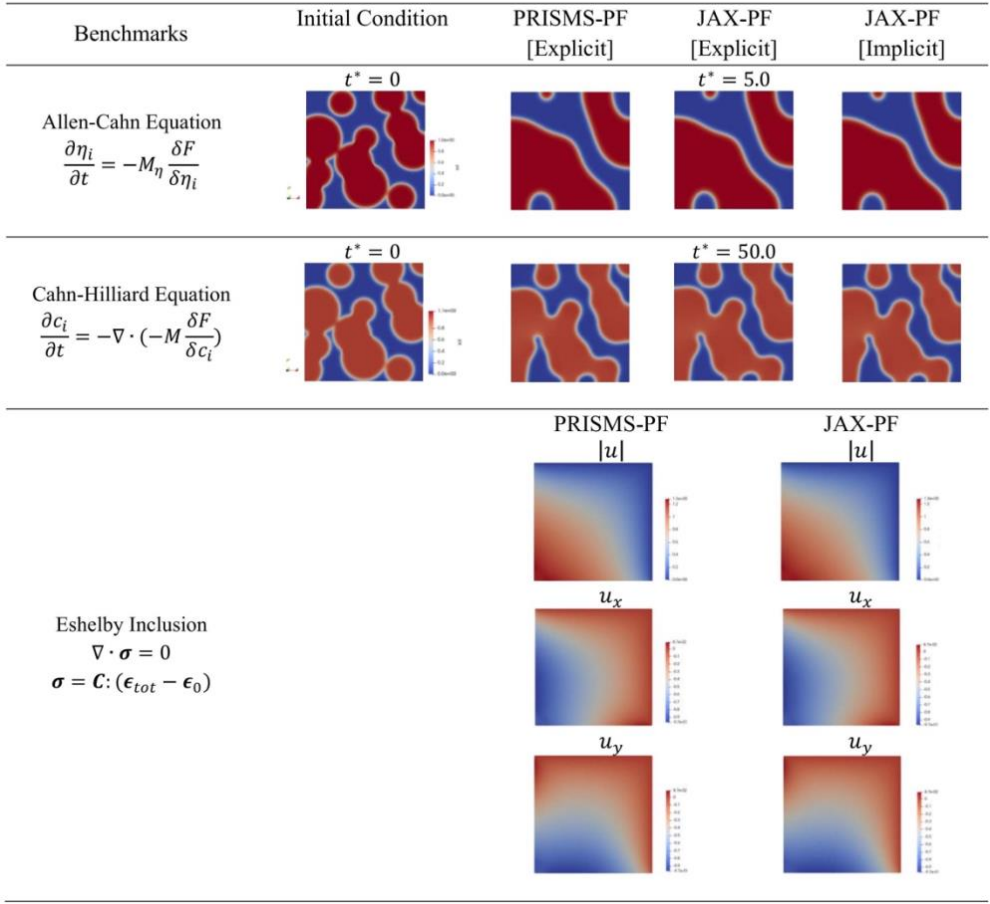

Figure 2. Validation of benchmark problems in JAX-PF, including the Allan-Cahn equation for non-conserved PF variable, the Cahn-Hilliard equation for conserved PF variable, and the Eshelby inclusion problem capturing displacement fields around a misfitting inclusion. PRISMS-PF is limited to explicit integration, while JAX-PF offers both explicit and implicit schemes, with results demonstrating consistency between the two frameworks.

### 3.2 Computational cost: JAX-PF vs. PRISMS-PF

Computational performance is a critical concern in PF modeling—not only for forward simulations but also for inverse design, which requires repeated forward solves of PF calculations during optimization. To evaluate performance, we conducted benchmark tests across varying numbers of field variables and mesh resolutions, using identical boundary conditions, as detailed in Supplemental Note 3. For a fair comparison, JAX-PF was run with an explicit time-stepping scheme on a GPU, while PRISMS-PF, one of the most efficient open-source PF solvers, was executed with explicit integration on CPUs using MPI parallelization across 24 cores. Wall time measurements relative to the number of degrees of freedom (DOF) are summarized in Figure 3.

For the Allen-Cahn equation with a single non-conserved PF order parameter, JAX-PF running 500,000 time steps on a GPU demonstrated clear advantage for medium to larger-size problems (meshes of $256^2$, $512^2$, $1024^2$, and $2048^2$), as shown in Figure 3a. For instance, with $2048^2$ mesh (~4,19 million DOF), JAX-PF (explicit) required 1,665 s (~28 min), whereas PRISMS-PF took 8,460 s (~2.35 hours). JAX-PF (explicit) achieved 5.1× speedup compared to PRISMS-PF with MPI. The hardware specifications used in these benchmarks are provided in Supplementary Note 4.

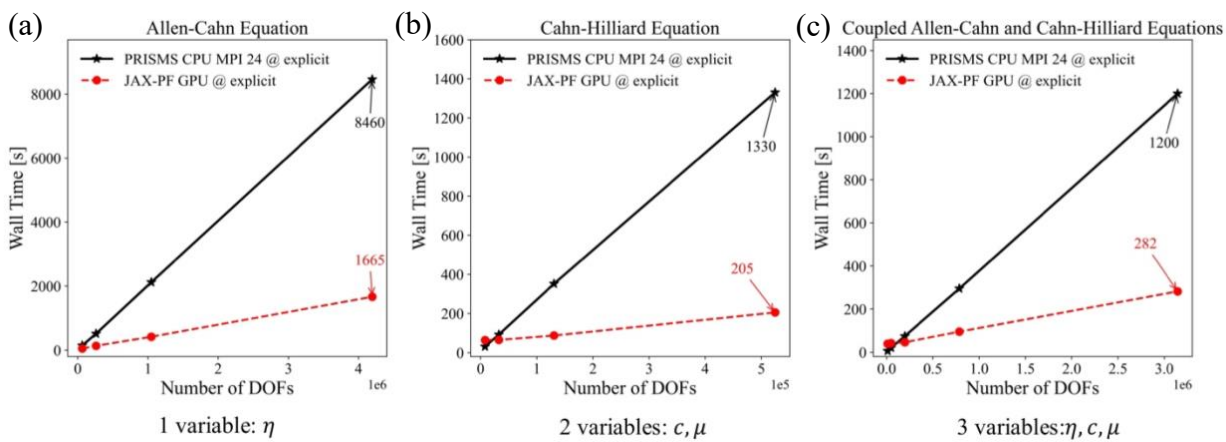

Figure 3. A summary of the performance of benchmarks using explicit solver with different levels of mesh resolution, including (a) Allen-Cahn equation consider one variable with 500,000 time steps ($256^2$, $512^2$, $1024^2$, and $2048^2$ mesh), (b) Cahn-Hilliard equation consider two variables with 500,000 steps ($64^2$, $128^2$, $256^2$, $512^2$ mesh), and (c) coupled Allen-Cahn and Cahn-Hilliard equations with three variables with 1500,000 steps ($64^2$, $128^2$, $256^2$, $512^2$, $1024^2$ mesh). Here, "PRISMS-PF CPU MPI 24" means PRISMS-PF runs with twenty-four processes of MPI parallel programming.

For the Cahn-Hilliard equation, a fourth-order partial difference equation, it involves two variables (concentration $c$ and chemical potential $\mu$), as detailed in the Section 2. PRISMS-PF with MPI performs competitively with smaller DOF ($64^2$ mesh). However, JAX-PF (explicit) on GPU exhibits better performance for larger meshes ($128^2$, $256^2$, $512^2$) for moderate to larger-size problems, as shown in Figure 3b. The same conclusion holds for the coupled Allen–Cahn and Cahn–Hilliard equation with three field variables, including $\eta, c$, and $\mu$, across mesh resolutions ranging from $64^2$ to $1024^2$.

These performance gains stem from `JAX`'s XLA backend, which generates highly optimized computational kernels, combined with just-in-time (JIT) compilation that accelerates vectorized operations. In contrast, MPI-based acceleration in `PRISMS-PF` is less efficient at very large DOF, due to communication overhead and memory constraints when handling transient global stiffness matrices. By leveraging GPU parallelism, `JAX-PF` achieves substantially improved scalability, with performance advantages particularly evident in nonlinear cases, which is consistent with observations in our `JAX-CPFEM` benchmarks [58].

Compared to explicit schemes, implicit solvers require significantly fewer time steps to achieve stable solutions, making them particularly attractive as a foundation for sensitivity analysis and inverse design. For a differentiable platform, this advantage is twofold. First, explicit schemes often demand an extremely large number of time steps due to stability constraints, leading to prohibitively long trajectories when simulating microstructure evolution at realistic scales. Such long trajectories not only increase computational cost but also pose severe challenges for AD, as reverse-mode AD requires either storing all intermediate states or repeatedly recomputing them during backpropagation. Second, implicit integration alleviates these issues by allowing much larger time increments, thereby shortening the trajectory length and reducing memory requirements, while still ensuring numerical stability and accuracy.

However, existing open-source frameworks provide limited support in this regard: `PRISMS-PF` implements only explicit schemes, while `MOOSE` offers implicit solvers but without GPU acceleration and differentiable capabilities, leading to higher computational cost. In contrast, `JAX-PF` provides both explicit and implicit solvers, combining the stability advantages of implicit integration with GPU acceleration and AD. Since computational performance remains critical for large-scale problems, we further examined the scalability of implicit solvers of `JAX-PF` across different mesh resolutions for the same benchmark cases. As shown in Figure 4a to c, the implicit solver scales well across different problem sizes (degrees of freedom), and it permits larger time steps with significantly fewer iterations compared to explicit solvers.

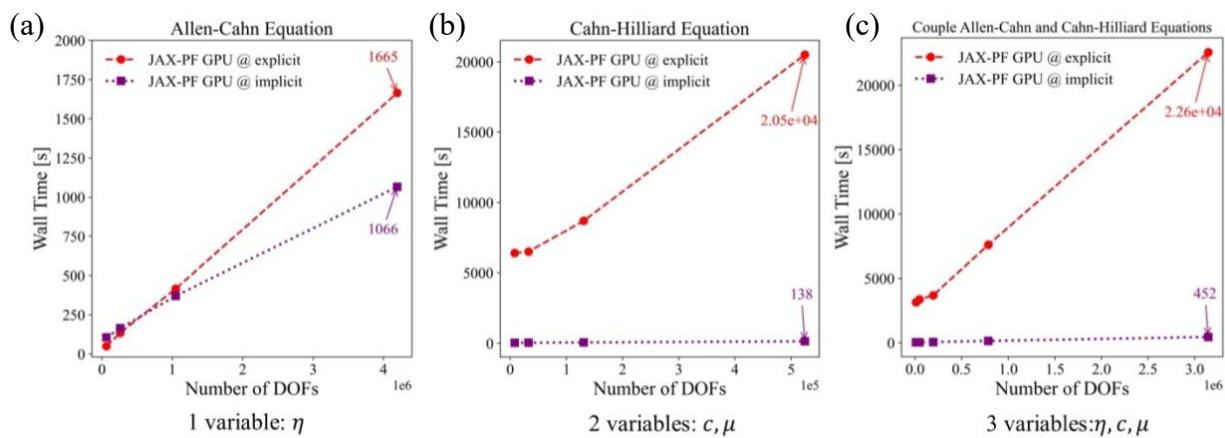

Figure 4. A summary of the performance of benchmarks using explicit and implicit solver with different levels of mesh resolution, including (a) Allen-Cahn equation consider one variable using explicit (500,000 time steps) and implicit (50 time steps) solvers ($256^2$, $512^2$, $1024^2$, and $2048^2$ mesh), (b) Cahn-Hilliard equation consider two variables using explicit (50,000,000 time steps) and implicit (50 time steps) solvers ($64^2$, $128^2$, $256^2$, $512^2$ mesh), and (c) coupled Allen-Cahn and Cahn-Hilliard equations with three variables using explicit (120,000,000 time steps) and implicit (50 time steps) solvers ($64^2$, $128^2$, $256^2$, $512^2$, $1024^2$ mesh).

It is important to emphasize that a direct runtime comparison between explicit and implicit solvers must be made with caution. A fair comparison requires enforcing the same accuracy criterion, which strongly depends on the admissible time step size. These time steps are not arbitrary as they are governed by the intrinsic properties of the underlying PDEs and stability conditions of the numerical scheme. Therefore, to ensure a consistent level of accuracy, we deliberately adopted smaller time step sizes for the explicit solver in these benchmark studies, as shown in Figure 4. In this work, we mainly emphasize the methodological differences between explicit and implicit approaches, rather than their absolute computational cost. This distinction highlights the advantage of implicit solvers in providing a practical foundation for sensitivity analysis and inverse design, where reducing the number of time steps is essential for efficiency.

### 3.3 AD-based Sensitivity Analysis of benchmarks

Sensitivities are indispensable for gradient-based optimization algorithms for inverse problems [46]. To illustrate the differentiability of `JAX-PF` with its implicit time-integration scheme, we conducted sensitivity analyses on three benchmark problems, including Allen–Cahn, Cahn–Hilliard, and Eshelby inclusion. These AD-based sensitivities enable calibration of material parameters for various kinds of phase transformations, such as martensite transformation and precipitation, featuring targeted microstructure through gradient-based optimization. Details of such inverse problems will be presented in the next section.

For Allen-Cahn and Cahn-Hilliard benchmarks, we considered a domain with size of 100 × 100 discretized with a 64 × 64 mesh. Sensitivity analyses focused on microstructural characteristics with respect to material parameters, including the mobility of interface ($M_\eta$, $M_c$) and the gradient energy coefficient $\kappa$, as described in the Section 2.1. The microstructure response, $\hat{O}(\boldsymbol{\theta})$, was defined as the spatial integral of PF variables over the domain: the order parameter $\eta$ in the Allen-Cahn case and concentration $c$ in the Cahn-Hilliard case. For Eshelby inclusion, we concern the quantitative effects of Young's modulus $E$ and Poisson's ratio $v$ to the total volume of the domain. Input parameters were collected and flattened into a vector $\boldsymbol{\theta} = [\theta_1, \theta_2]$, and the microstructure response and its corresponding sensitivity were then expressed as follows:

$$\hat{O}_{Allen-Cahn}(\boldsymbol{\theta}) = \frac{1}{V} \int \eta(M_\eta, \kappa) \, d\Omega = \frac{1}{V} \int \eta(\boldsymbol{\theta}) \, d\Omega, \tag{19}$$

$$\hat{O}_{Cahn-Hilliard}(\boldsymbol{\theta}) = \frac{1}{V} \int c(M_c, \kappa) \, d\Omega = \frac{1}{V} \int c(\boldsymbol{\theta}) \, d\Omega, \tag{20}$$

$$\hat{O}_{Inclusion}(\boldsymbol{\theta}) = V(E, v) = V(\boldsymbol{\theta}), \tag{21}$$

$$\boldsymbol{J} = \frac{\partial \hat{O}(\boldsymbol{\theta})}{\partial \boldsymbol{\theta}}. \tag{22}$$

This sensitivity analysis provides insight into how small variations in material coefficients influence the resulting microstructure. Additionally, it enables automatic calibration of PF models through gradient-based optimization using experimental data, such as computed tomography (CT) [79], 3D X-ray diffraction [80], or diffraction contrast tomography (DCT) [81]. In this context, the derivative of the objective function $\hat{O}(\boldsymbol{\theta})$ —for example, the similarity between experimental imaging and simulation—with respect to material parameters can, in principle, be derived through the chain rule (see Supplementary Note 2). However, because PF models involve strongly nonlinear couplings among multiple field variables, deriving analytical sensitivities is challenging, time-consuming, and error prone.

A common alternative is finite-difference-based numerical differentiation (FDM), as detailed in Supplementary Note 6, which approximates derivatives by perturbing each parameter and re-running the forward simulation [78]. While conceptually simple, this approach is computationally expensive for PF models, since each component of the derivative requires multiple forward solves, and the cost scales with the number of parameters. Moreover, FDM approximations may introduce errors, which can be problematic in quantitative PF studies requiring precise sensitivities, and lead to incorrect conclusions. By contrast, `JAX-PF` overcomes these issues by directly using AD with adjoint-based methods, producing exact gradients (up to machine precision) in a fully automated manner. This makes the approach both efficient and user-friendly.

Table 1. Comparison between AD and FDM sensitivity calculations for each benchmark with parameter vector $\boldsymbol{\theta} = [\theta_1, \theta_2]$

|  |  | $\dfrac{\partial \hat{O}(\boldsymbol{\theta})}{\partial \theta_1}$ | $\dfrac{\partial \hat{O}(\boldsymbol{\theta})}{\partial \theta_2}$ |
|---|---|---|---|
| Allen-Cahn | AD results | 102.502519 | 72.341267 |
| Allen-Cahn | FDM results | 102.512117 | 72.340078 |
| Allen-Cahn | Difference | 0.0676‰ | 0.0164‰ |
| Cahn-Hilliard | AD results | 2.735689 | 1.549975 |
| Cahn-Hilliard | FDM results | 2.735787 | 1.549971 |
| Cahn-Hilliard | Difference | 0.0358‰ | 0.00258‰ |
| Eshelby inclusion | AD results | 12.447285 | -6.223642 |
| Eshelby inclusion | FDM results | 12.447379 | -6.223727 |
| Eshelby inclusion | Difference | 0.00755‰ | 0.0137‰ |

Note: Non-dimensional treatment was applied for units; Difference $= \dfrac{|\text{AD results} - \text{FDM results}|}{\text{AD results}}$

For validation, Table 1 compares sensitivities obtained from AD and FDM for the three benchmarks. The results show less than 0.1‰ difference between the two methods, confirming the correctness and reliability of `JAX-PF`'s differentiability, even for highly nonlinear problems. Furthermore, in multi-PF and multi-physics problems (e.g., shape-memory alloys combining Allen–Cahn dynamics with mechanical fields), the number of material parameters often exceeds the number of objective functions, which would require repeated forward PF runs under FDM. In contrast, AD in `JAX-PF` produces the full sensitivity vector in a single backward pass, yielding substantial acceleration, especially when combined with GPU execution.

## 3.4 Applications

The flexibility of `JAX-PF` allows it to be applied across a wide range of scenarios, such as the evolution of morphology, grain size, orientation distribution, transition path, defect structure, and stress-strain distribution of different phases under complex internal and external effects. Building on the well-established differentiable benchmark models introduced earlier, in this section, we demonstrate five representative applications that span from solidification to solid-state transformations: grain growth, static recrystallization, spinodal decomposition, precipitation with the WBM model, and precipitation with the KKS model. These models are established on the open-source library from `PRISMS-PF` [56] and solid foundation of conventional PF, highlighting the versatility of the framework in capturing diverse microstructural processes. Simulation results are presented in Figure 5, and the PF frameworks and their corresponding simulation details for each case are summarized in Table 2. More details about our `JAX-PF`'s features are introduced in Table 2.

First, the framework of `JAX-PF` provides flexibility by offering both explicit and implicit schemes for all implemented models. Explicit schemes are relatively simple to implement, computationally inexpensive per time step, and well-suited for problems involving rapid transients. However, they require very small time-steps, limited by the Courant-Friedrichs-Lewy condition, and it is only first-order accurate. In contrast, implicit schemes are unconditionally stable and allow much larger time increments, which drastically reduces the total number of steps required. Their main drawback lies in the need to solve a nonlinear system at each step, which is more computationally demanding per iteration. This trade-off means that explicit methods are often preferred for short, transient dynamics, while implicit methods excel in simulations that span long physical times or involve strongly coupled fields. The dual capability of both explicit and implicit solvers makes `JAX-PF` adaptable allows users to select the most suitable scheme for their specific problem. Moreover, by directly comparing explicit and implicit implementations of `JAX-PF` against `PRISMS-PF` benchmarks, we confirmed both the accuracy and efficiency of our framework, thereby validating its reliability for diverse PF applications.

Table 2. Overview of five PF cases and key capabilities demonstrated by the proposed pipeline.

| | Applications | Control Equation(s) | Variables | Reference |
|---|---|---|---|---|
| Case Study 1 | Grain Growth | Allen-Cahn | Order Parameter $\eta$ | [56, 82] |
| Case Study 2 | Spinodal Decomposition | Cahn-Hilliard | Concentration $c$ | [56] |
| Case Study 3 | Precipitation (WBM model) | Allen-Cahn<br>Cahn-Hilliard<br>Momentum Balance | Order Parameter $\eta_1, \eta_2, \eta_3$<br>Concentration $c$<br>Displacement $\boldsymbol{u}$ | [56, 83] |
| Case Study 4 | Precipitation (KKS model) | Allen-Cahn<br>Cahn-Hilliard<br>Momentum Balance | Order Parameter $\eta$<br>Concentration $c$<br>Displacement $\boldsymbol{u}$ | [56, 83] |
| Case Study 5 | Static Recrystallization | Allen-Cahn | Order Parameter $\eta$ | [14, 82] |

Second, `JAX-PF` leverages AD to relieve users from manually deriving derivatives of different terms, such as free-energy functionals, which are often complex and case-specific. This capability is particularly advantageous for advanced models such as the KKS formulation [40], where additional phase concentrations must be introduced and the resulting free-energy expressions become cumbersome. In most of popular PF software, these derivatives must be carefully derived and coded by hand, a process that is error-prone and limits extensibility. With `JAX-PF`, AD ensures machine-precision derivatives, enabling researchers to focus on model development and physical insight rather than on mathematical bookkeeping.

Third, on top of `JAX-FEM`, the architecture of `JAX-PF` is fully vectorized at the finite element cell level, and we extend this vectorization to multiple PF variables described with similar formulations. Instead of relying on explicit for-loops, as is common in `Fortran`, `Matlab`, or `C/C++` implementation, `JAX-PF` employs `jax.vmap`, a core function of `JAX`, for vectorized operations across variables. In grain growth simulations, for instance, each grain can be treated as a separate order parameter, and the vectorized formulation updates all grains on each quadrature point simultaneously using the similar Allen–Cahn equations with different material parameters and free-energy expression. This approach maximizes GPU utilization and achieves highly efficient parallel scaling with respect to the number of PF variables. As an example, in the grain growth case study shown in Figure 5a, simulations with six to twenty order parameters required only ~270 seconds on a GPU using the implicit solver. Such performance makes it practical to investigate microstructure evolution in polycrystals with far larger numbers of grains than previously feasible, opening the door to ultra-large-scale PF studies that were once constrained by applying different weak form formulations for each individual PF variable.

Fourth, nucleation criteria can be incorporated directly into the `JAX-PF` framework [37], making it possible to simulate nucleation-driven processes, such as static recrystallization, as shown in Figure 5e. By introducing probabilistic nucleation models within the PF formulation, which is related to the dislocation field of each grain/grain boundary, `JAX-PF` enables the study of nucleation density, spatial distribution, and orientation selection, which are critical for understanding microstructural refinement in thermomechanical processing [84, 85]. This functionality bridges the gap between mesoscale PF modeling and realistic microstructure design strategies.

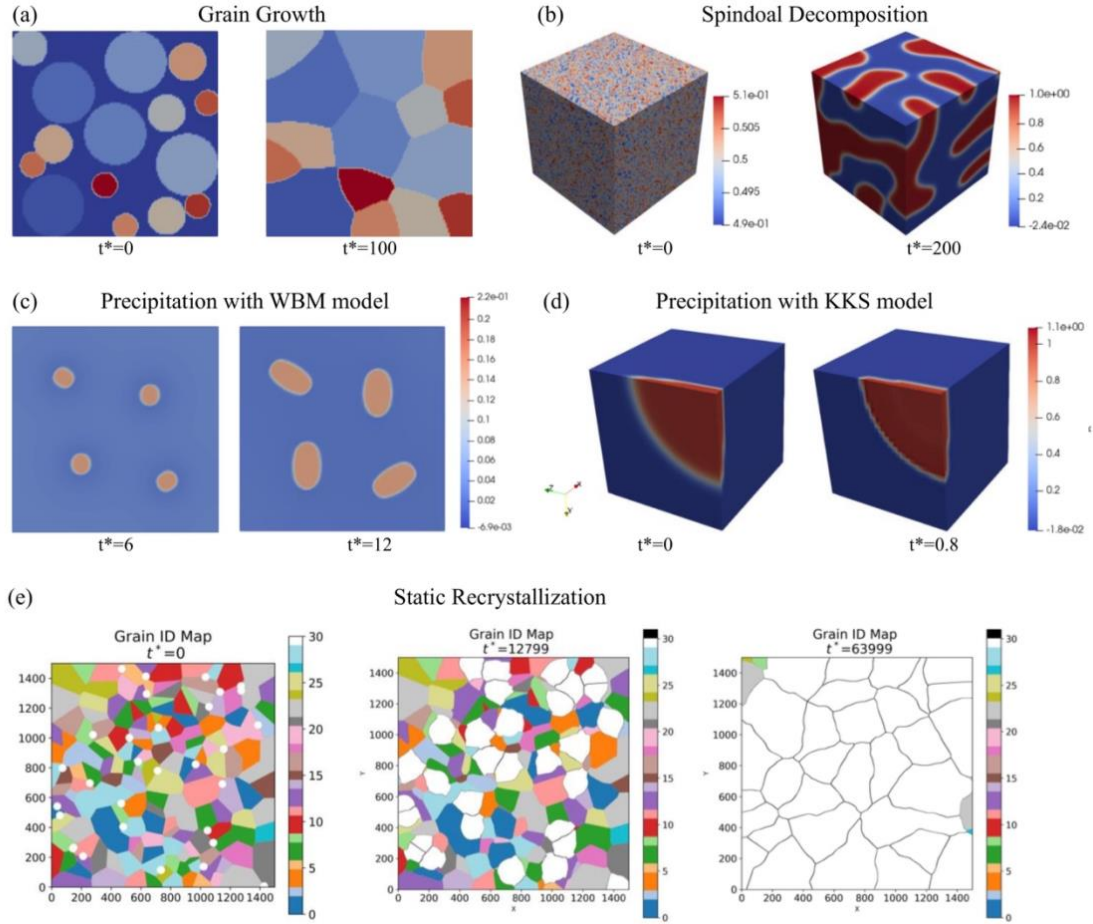

Figure 5. Examples of five applications in `JAX-PF` utilizing both explicit and implicit time-integration schemes. (a) shows a 2D grain growth case, showing the microstructure at the initial state (left) and after evolution (right). Grain colors indicate identification numbers, which remain consistent as the order parameters are reassigned during evolution. (b) illustrates the composition field during spinodal decomposition, evolving from initial fluctuations (left) to the formation of two distinct phases (right). Subfigure (c) and (d) present evolution of precipitation process in an Mg-Nd alloy Wheeler-Boettinger-McFadden (WBM) and Kim-Kim-Suzuki (KKS) models, respectively. A comparison of visualization of the PF variables between `JAX-PF` (implicit), `JAX-PF` (explicit), and `PRISMS-PF` (explicit), whose applications were released open-source, has been conducted and validated the accuracy of our software. (e) shows simulated recrystallized microstructure after times of t* = 0, 12799, 63999. White color represents the new recrystallized grains, and black color represents grain boundaries.

Finally, built on the same `JAX-FEM` platform, `JAX-PF` and `JAX-CPFEM` [58, 86] share the same ecosystem and can be combined into a single GPU-accelerated differentiable framework. This integration supports multiscale PF–CPFEM simulations of process–structure–property relationships under complex fields [63, 87, 88], such as coupled thermo-mechanical phenomena. For instance, as shown in Figure 6a, `JAX-PF` captures the solidification process and the distribution of 15 grains, while `JAX-CPFEM` predicts the corresponding mechanical response (Figure 6d) under quasi-static tensile loading up to 0.5% over 50 steps, as shown in Figure 6c. The realization detail of CPFEM simulation is detailed in Supplementary Note 5. The resulting stress fields and deformation mechanisms are directly linked to the evolving microstructure, as shown in Figure 6b. This pipeline can be extended to other scenarios, such as dynamic recrystallization, by combining PF for grain nucleation and growth with CPFEM for linking material microstructure deformation mechanisms to resulting mechanical properties. Crucially, the differentiable nature of the framework ensures that such coupled simulations can be directly embedded into gradient-based optimization pipelines. Importantly, this capability aligns directly with the objectives of the HAMMER ERC project on Hybrid Autonomous Manufacturing, where digital twins and adaptive control strategies require integrated models that connect processing routes to performance [4]. By enabling property-targeted microstructure design within a differentiable and GPU-accelerated ecosystem, the coupling of `JAX-PF` and `JAX-CPFEM` provides a powerful computational foundation for co-designing of material and manufacturing processes.

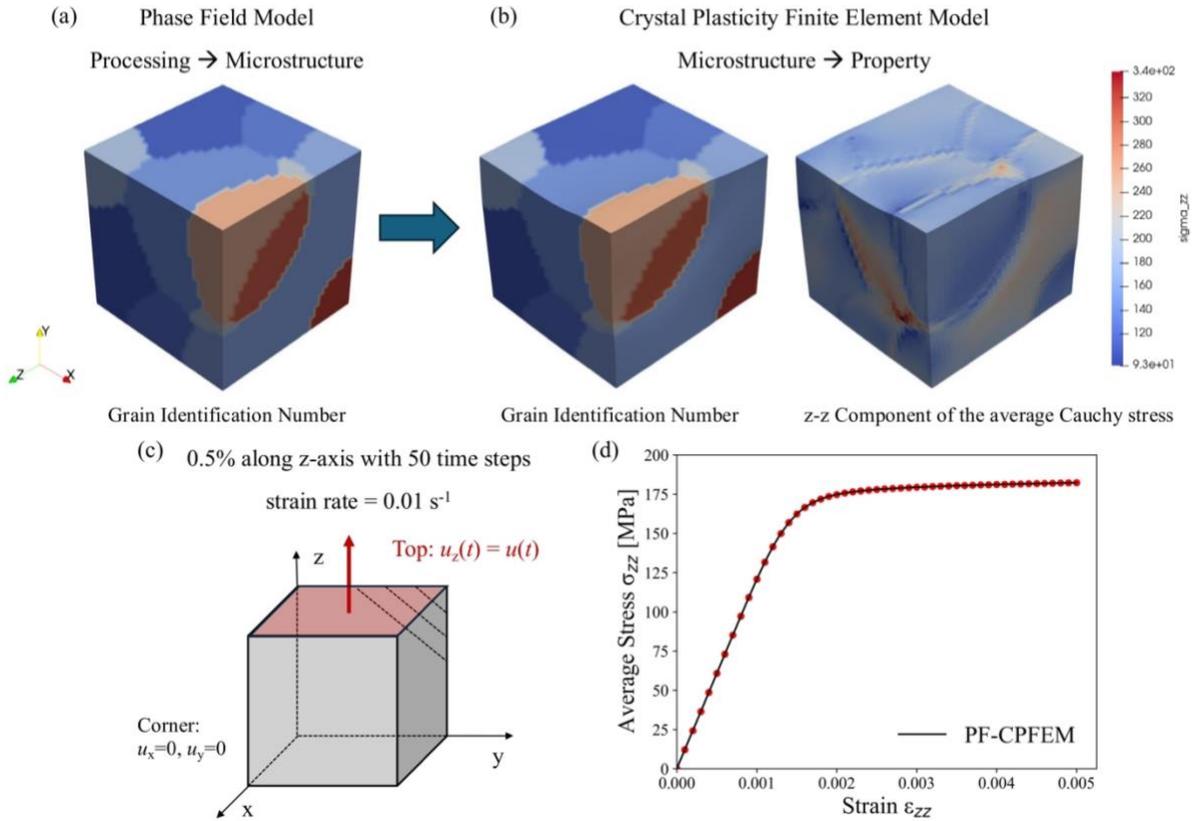

Figure 6. Integrated `JAX-PF` and `JAX-CPFEM` framework for process–structure–property relationship. Subfigure (a) shows the grain distribution obtained from `JAX-PF` solidification simulation, while Subfigure (b) shows the local microstructure information of polycrystal copper, including grain distribution and the z-z component of Cauchy stress $\sigma_{zz}$, mapped to deformed sample (scale factor: 20) predicted by `JAX-CPFEM`, under the (c) boundary condition: bottom and one corner fixed; top surface loaded by prescribed displacement. Subfigure (d) illustrates the average stress $\sigma_{zz}$ with respect to the applied strain $\varepsilon_{zz}$ under quasi-static loading up to 0.5% strain.

## 3.5 Inverse Problem: Calibration of material parameters using differentiable phase-field simulations

The establishment of reliable and comprehensive materials database is a central goal of the Materials Genome Initiative (MGI) [89, 90] and a key component of the Integrated Computational Materials Engineering (ICME). The accuracy of such database depends critically on accurate predictive simulations of microstructure evolution, where quantitative PF models play a central role by linking processing conditions to microstructural features. The predictive capability of quantitative PF models, however, rely heavily on two factors [26]: (i) accurate representation of the underlying physical mechanisms, such as free-energy formulations and interfacial energies, and (ii) reliable determination of material parameters. While substantial progress has been made on the first point, the determination of material parameters remains a significant challenge. Many studies rely on qualitative comparisons between experiments and simulations, or on averaged metrics such as mean particle size or interface area per volume. These approaches often introduce large uncertainties, leaving the quantitative calibration of PF parameters as an open and pressing problem [91-94].

To address this challenge, researchers have pursued two main approaches. Multiscale modeling methods (first-principles [95, 96], molecular dynamics [97, 98], Monte Carlo [99, 100]) can provide estimates of free energies or diffusion coefficients, but their complexity, uncertainty in bridging scales, and the need for experimental validation limit their utility. Alternatively, experimental image-based calibration seeks to fit PF simulations to microstructures captured by advanced microscopy techniques, such as Scanning Electron Microscope (SEM) [101], Transmission Electron Microscopy (TEM) [102], and 3D X-Ray Diffraction (3DXRD) [80]. With the aid of image segmentation algorithms [103-106], grain boundaries, particle detection, and different phases can be extracted, providing rich reference data. Then, based on the segmented images, this approach is often carried out in a trial-and-error manner, which is computationally expensive given that large-scale PF simulations may require days of wall time.

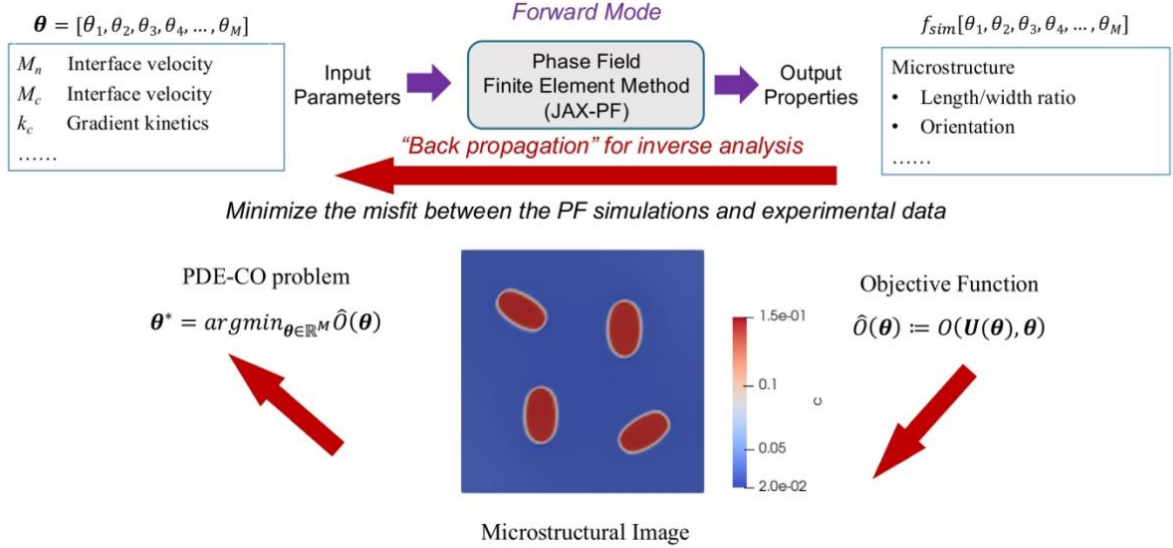

Figure 7. Workflow of the inverse problem for calibration of PF material parameters utilizing differentiable `JAX-PF`. The objective function, to be minimized, is modeled as the misfit between the PF simulations and segmented images from microscopic experimental data, such as Scanning Electron Microscope (SEM), Transmission Electron Microscopy (TEM), and 3D X-Ray Diffraction (3DXRD).

To address these challenges, a more systematic solution lies in inverse problem-solving techniques, where experimental observations are treated as inputs to identify high-dimensional model parameters, as shown in Figure 7. The overall workflow follows a gradient-based optimization strategy, where search directions are guided by the gradient of the objective at the current parameter estimate. And the central requirement is the accurate and efficient computation of sensitivities—that is, the gradients of objective functions with respect to design variables. Accurate evaluation of sensitivities—defined as the gradients of objective functions with respect to design parameters—is essential for quantifying how material, microstructural, and processing parameters influence microstructural characteristics such as phase distributions, volume fractions, and morphologies. In this study, we introduce the concept of differentiable PF modeling, in which arbitrary end-to-end derivatives are made available within the proposed `JAX-PF` pipeline, as defined in Eq. (18). This capability enables the direct application of gradient-based optimization methods to solve inverse problems such as parameter calibration. To the best of our knowledge, this study represents the first application of AD-based sensitivities to PF-based material parameter calibration, thereby opening the door to efficient, gradient-based solutions of inverse problems across a wide range of applications.

In this section, we consider the inverse calibration of material parameters in a Mg–Nd alloy using synthetic microstructural imaging. A synthetic two-dimensional phase microstructural image (bottom in Figure 7) of multiple variants was created based on the reference parameters summarized in Table 3, and serves as the ground truth for calibration. We focus on the calibration of eight PF parameters, denoted as a vector $\boldsymbol{\theta} = [E_\alpha, E_\beta, \mu_\alpha, \mu_\beta, M_c, M_{\eta_1}, M_{\eta_2}, M_{\eta_3}] = [\theta_1, \theta_2, \ldots, \theta_8]$, which includes both kinetic coefficients and mechanical properties. The targeted microstructural features, $y_{i,j}$, represent the existence of different variants $j$ ($j$=1, 2, 3) at mesh point $i$. In summary, the calibration is formulated as a PDE-CO problem, where the objective function $\hat{O}(\boldsymbol{\theta})$ measures the difference between the reference and simulated microstructures:

$$\hat{O}(\boldsymbol{\theta}) = \sum_{i=1}^{n} \sum_{j=1}^{m} w_j \times (y_{i,j} - f_{sim,i,j}(\boldsymbol{\theta}))^2, \quad (23)$$

where $f_{sim,i,j}(\boldsymbol{\theta})$ denotes the PF simulation (sim) results of the microstructure at the final stages predicted by `JAX-PF` for mesh point $i$ and $j$-th variant, and $w_j$ is the weight assigned to each variant. It is worth noting that the sensitivity analysis and subsequent gradient-based optimization do not dependent on the specific form of the objective function, as long as it is smooth and differentiable. Based on the chain rule and the implicit PF scheme, the required sensitivities $\frac{\partial \hat{O}(\boldsymbol{\theta})}{\partial \boldsymbol{\theta}}$ were efficiently computed using the AD function in `JAX` called '`jax.vjp`', which stands for vector-Jacobian product. This function eliminates the need for tedious and error-prone manual derivations and has been validated across multiple benchmarks presented earlier. And for the gradient-based optimization task, we employed the limited-memory BFGS (L-BFGS) algorithm [107], implemented in the `SciPy` package [108], which is well-suited for large-scale optimization

Table 3. Material parameters used to generate reference microstructure image for the Mg-Nd alloy, along with their optimization bounds. We decomposed the input parameters $\boldsymbol{\theta} = [\theta_1, \theta_2, ..., \theta_8]$ into two parts, including an initial estimate $\boldsymbol{x} = [x_1, x_2, ..., x_8]$, reflecting the approximate value of each material parameter based on their physical meanings, and a set of scaling coefficients $\boldsymbol{a} = [a_1, a_2, ..., a_8]$.

| Materials Parameters | Physical Meaning | Synthetic target $\theta_i = a_i \times x_i$ | Calibrated results $\theta_{i,cal} = a_{i,cal} \times x_i$ |
|---|---|---|---|
| $E_\alpha$ | Young's modulus of matrix | $1.0 \times 40.0$ | $0.959534 \times 40.0$ |
| $E_\beta$ | Young's modulus of $\beta$-phase | $1.0 \times 50.0$ | $0.731421 \times 50.0$ |
| $\mu_\alpha$ | Poisson's ratio of matrix | $1.0 \times 0.3$ | $0.961225 \times 0.3$ |
| $\mu_\beta$ | Poisson's ratio of $\beta$-phase | $1.0 \times 0.3$ | $0.960839 \times 0.3$ |
| $M_c$ | Mobility value for the concentration field $c$ | $1.0 \times 1.0$ | $0.733411 \times 1.0$ |
| $M_{\eta_1}$ | Mobility value for the structural order parameter field $\eta_1$ | $1.0 \times 100.0$ | $0.958086 \times 100.0$ |
| $M_{\eta_2}$ | Mobility value for the structural order parameter field $\eta_2$ | $1.0 \times 100.0$ | $0.96427 \times 100.0$ |
| $M_{\eta_3}$ | Mobility value for the structural order parameter field $\eta_3$ | $1.0 \times 100.0$ | $0.728702 \times 100.0$ |

Note: Non-dimensional treatment was applied for units

To solver high-dimensional optimization efficiently, we decomposed the input parameters $\boldsymbol{\theta} = [\theta_1, \theta_2, ..., \theta_8]$ into two parts, including an initial estimate $\boldsymbol{x} = [x_1, x_2, ..., x_8]$, reflecting the approximate value of each material parameter based on their physical meanings, and a set of scaling coefficients $\boldsymbol{a} = [a_1, a_2, ..., a_8]$. This approach allows the model to balance conflicting input variables more effectively by assigning appropriate weights to each parameter according to its significance for each goal, see our recent paper for implementation from the programming side in details [86]. Upper ($a_{i,upper} = 1.5$) and lower bounds ($a_{i,lower} = 0.7$) were set based on physical meaning and previous calibration experience for each calibrated parameter during the optimization process. The optimization begins with an initial guess far from the ground truth, as indicated by the first point in Figure 8a. Each iteration corresponds to a gradient query, and the objective function (defined in Eq. (23)) decreases rapidly with the AD-based derivatives. Within only 18 iterations, the percentage reduction in the objective function value fell below 5%. Using the calibrated parameters obtained at this stage, as summarized in Figure 8b, we conducted a sanity test by running forward `JAX-PF` simulations. As shown in Figure 8c, the calibrated microstructure (blue dots) closely matches the reference morphology of different variants (red dots), which are represented by the concentration ranging from 0.12 to 0.16, validating the effectiveness of our AD-based calibration pipeline for the high-dimensional inverse problem.

Despite these encouraging results, as shown in Figure 8c, several points need further discussion. First, three of the calibrated parameters, including mobility value for the structural order parameter field $\eta_3$, mobility value for the concentration field $c$, and Poisson's ratio of matrix $\mu_\alpha$, show distinct deviations from the reference data compared to others. This result reflects the ill-posed nature of parameter calibration, where multiple solutions may exist. Our framework has the potential to incorporate various regularization techniques (e.g., Tikhonov regularization [109], constraint enforcement, or smoothing [110]) to stabilize the problem and make this inverse problem well-posed. Second, the current objective function relies solely on morphological comparison, whereas PF represents a multi-physics process influenced by factors such as composition diffusion. As shown in Figure 8e and Figure 8f, differences remain in the concentration distribution between the reference and calibrated results. These results highlight the need for objective functions that integrate additional experimental data and richer physical information, such as average concentration, phase-specific concentrations, or complementary data from SEM, TEM, or HAADF-STEM. These directions form the basis of our future research.

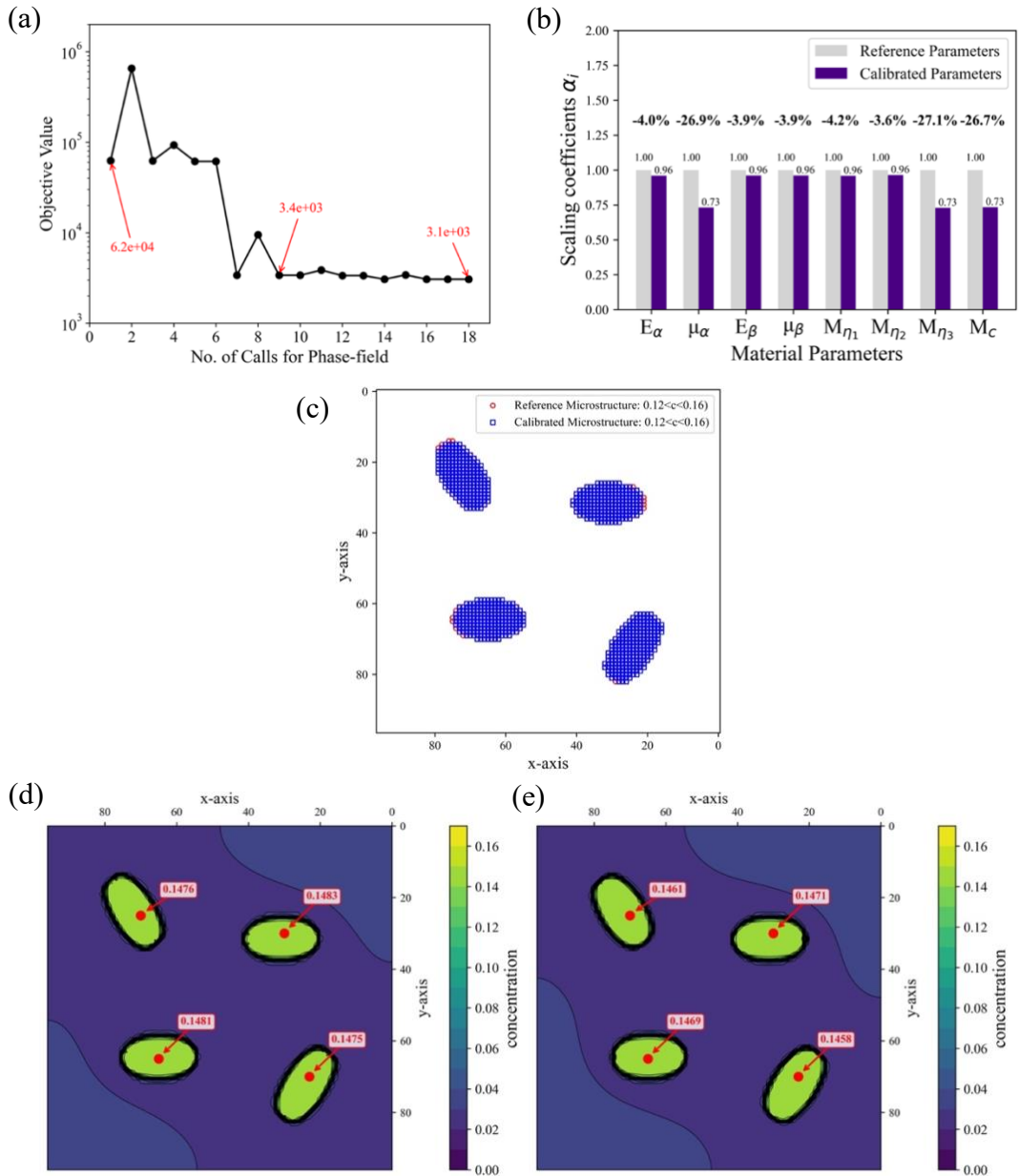

Figure 8. Calibration of material parameters for multi-variants in Mg-Nd alloys using differentiable PF simulations. (a) Evolution of the objective function over optimization iterations, based on synthetic microstructural imaging used as ground truth (bottom subfigure of Figure 7). (b) Comparison between the reference parameters and calibrated parameters. For clearly, the calibrated parameters $\boldsymbol{\theta} = [\theta_1, \theta_2, ..., \theta_8]$ are shown as a set of scaling coefficients $\boldsymbol{a} = [a_1, a_2, ..., a_M]$ relative to the reference values $\boldsymbol{x} = [x_1, x_2, ..., x_8]$. (c) Comparison of the reference (red) and calibrated (blue) microstructural morphologies of different variants, where precipitate variants are identified by concentrations between 0.12 and 0.16. (d, e) Concentration distributions across the domain for the reference and calibrated cases, respectively.

## 4. CONCLUSIONS

In this work, we introduced `JAX-PF`, an open-source, GPU-accelerated, ease-of-use, and differentiable phase-field (PF) software package. Compared with existing open-source PF packages, `JAX-PF` stands out for its affordability, flexibility, and suitability for inverse design, making it accessible to a broad community of users. A key advantage is its support for both explicit and implicit solvers, enabling stable and efficient simulations across a wide range of PF applications. By leveraging automatic differentiation (AD), `JAX-PF` eliminates the need for manual derivation of Jacobians and free-energy derivatives, an especially critical feature for complex multi-variable and multi-physics models such as the KKS formulation. This flexibility allows users to readily incorporate new physics into PF models without tedious re-derivations. Validation of `JAX-PF` against standard benchmarks in `PRISMS-PF`, including Allen–Cahn, Cahn–Hilliard, coupled Allen–Cahn and Cahn–Hilliard, and Eshelby inclusion problems, confirmed the accuracy of `JAX-PF` and extended its capabilities beyond those of `PRISMS-PF` by incorporating both explicit and implicit solvers. Some conclusions can be drawn as follows.

1. We assessed computational performance relative to `PRISMS-PF`. For explicit scheming on a single GPU, `JAX-PF` achieved ~5× acceleration for problems exceeding four million DOF compared to `PRISMS-PF` on CPUs with MPI (24 cores). The implicit solver of `JAX-PF` further demonstrated strong scalability across varying DOF levels and numbers of variables, providing stable solutions with far fewer time steps than explicit schemes. These features of efficiency and fewer time steps are crucial for inverse problems requiring AD-based sensitivity analysis. The performance gains arise from GPU acceleration through the XLA backend and vectorized operations via `jax.vmap`, which enable simultaneous updates of multiple PF variables. In grain growth simulations, for example, each grain can be represented as a separate order parameter and updated in parallel. These capabilities allow JAX-PF to tackle ultra-large-scale simulations that were previously constrained by mathematical complexity and implementation overhead.

2. Beyond forward modeling, `JAX-PF` demonstrated robust differentiable capability for sensitivity analysis and inverse design. AD-based sensitivities matched FDM-based derivatives within 0.1‰ difference while providing far greater efficiency, particularly for high-dimensional parameter spaces. As a demonstration, we performed parameter calibration for multi-variant precipitation in Mg–Nd alloys. `JAX-PF` successfully reproduced experimental morphologies through AD-based optimization, addressing a problem long recognized as highly nonlinear, high-dimensional, and computationally intensive. To the best of our knowledge, very few studies [91-94] have tackled PF calibration at this level of complexity. This case highlights the potential of `JAX-PF` to serve as a foundation for high-dimensional inverse design pipelines, where processing conditions or free-energy coefficients can be optimized directly against experimental imaging data.

3. Because `JAX-PF` and `JAX-CPFEM` share the same `JAX-FEM` foundation, the two frameworks can be tightly coupled to realize multiscale simulations directly linking process–structure–property relationships. For instance, PF can be used to model grain nucleation and growth, while CPFEM captures the associated mechanical response, all within a single GPU-accelerated and differentiable framework. Such integration provides a powerful computational infrastructure for co-designing of material and manufacturing processes. In particular, it directly aligns with the objectives of initiatives like the HAMMER ERC project, where property-targeted microstructure design and hybrid autonomous manufacturing are central goals.

**Declaration of Competing Interest**

The authors declare that they have no known competing financial interests or personal relationships that could have appeared to influence the work reported in this paper.

**Data Availability**

All data generated or analyzed during this study are included in this published article.

**Code Availability**

`JAX-PF` was constructed on top of our recent work `JAX-FEM`. They can be freely downloaded from the following link: https://github.com/SuperkakaSCU/JAX-PF. For `JAX-CPFEM` mentioned in the paper, our recently developed differentiable crystal plasticity finite element package, it can be freely downloaded from the following link: https://github.com/SuperkakaSCU/JAX-CPFEM.

**Author Contributions**

Fanglei Hu: Conceptualization, Methodology, Software, Investigation, Writing - Original Draft; Jiachen Guo: Validation, Writing - Review & Editing; Stephen Niezgoda: Validation, Writing - Review & Editing; Wing Kam Liu: Writing - Review & Editing; Jian Cao: Conceptualization, Methodology, Resources, Supervision, Writing - Review & Editing.

**Acknowledgements**

The authors acknowledge helpful discussions with Dr. Jin Zhang, Dr. Deepak Sharma, and Dr. Stephen DeWitt.

**Funding**

The authors would like to acknowledge support from the Department of Defense Vannevar Bush Faculty Fellowship, USA N00014-19-1-2642, and from the NSF Engineering Research Center for Hybrid Autonomous Manufacturing Moving from Evolution to Revolution (ERC-HAMMER) under Award Number EEC-2133630.
36

**Fanglei Hu[a], Jiachen Guo[b,c], Stephen Niezgoda[d], Wing Kam Liu[b,c], Jian Cao[a*]**

[a] Department of Mechanical Engineering, Northwestern University, Evanston, IL 60208, USA

[b] Theoretical and Applied Mechanics, Northwestern University, Evanston, IL, 60208, USA

[c] HIDENN-AI, LLC, 1801 Maple Ave, Evanston, 60201, IL, USA

[d] Department of Materials Science and Engineering, The Ohio State University, Columbus, OH 43210, USA

**Supplementary Note 1: Notation**

As a general scheme of notation, vectors are written as boldface lowercase letters (e.g. $\boldsymbol{a}$, $\boldsymbol{b}$), a second-order tensors as boldface capital letters (e.g. $\boldsymbol{A}$, $\boldsymbol{B}$), and fourth-order tensors as blackboard-bold capital letters (e.g. $\mathbb{C}$). For vectors and tensors, Cartesian components are denoted as, $a_i$, $A_j$, and $\mathbb{C}_{ijkl}$. All inner products are indicated by a single dot between the tensorial quantities of the same order, e.g., $\boldsymbol{a} \cdot \boldsymbol{b}$ ($a_i b_i$) for vectors and $\boldsymbol{A} \cdot \boldsymbol{B}$ ($A_{ij} B_{ij}$) for second-order tensors. The transpose, $\boldsymbol{A}^{\mathrm{T}}$, a tensor is denoted by a superscript "T", and the inverse, $\boldsymbol{A}^{-1}$, by a superscript "-1"

**Supplementary Note 2: Chain Rules for Derivative**

To derive the total derivative of $\hat{O}(\boldsymbol{\theta})$, mentioned in Eq. (18), with respect to parameters $\boldsymbol{\theta}$, we need to use chain rules as follows

$$\frac{d\hat{O}}{d\boldsymbol{\theta}} = \frac{\partial O}{\partial \boldsymbol{U}} \frac{d\boldsymbol{U}}{d\boldsymbol{\theta}} + \frac{\partial O}{\partial \boldsymbol{\theta}}, \tag{II.1}$$

where $\frac{d\boldsymbol{U}}{d\boldsymbol{\theta}}$ is justified by the implicit function theorem under certain mild conditions. Based on the constrain condition with respect to parameters $\boldsymbol{\theta}$,

$$\frac{d\boldsymbol{C}}{d\boldsymbol{\theta}} = \frac{\partial \boldsymbol{C}}{\partial \boldsymbol{U}} \frac{d\boldsymbol{U}}{d\boldsymbol{\theta}} + \frac{\partial \boldsymbol{C}}{\partial \boldsymbol{\theta}} = 0, \tag{II.2}$$

we can get the formulation of $\frac{d\boldsymbol{U}}{d\boldsymbol{\theta}}$ as follows

$$\frac{d\boldsymbol{U}}{d\boldsymbol{\theta}} = -\left(\frac{\partial \boldsymbol{C}}{\partial \boldsymbol{U}}\right)^{-1} \frac{\partial \boldsymbol{C}}{\partial \boldsymbol{\theta}}. \tag{II.3}$$

Thus, we can reformulate the total derivative of $\hat{O}(\boldsymbol{\theta})$ as follows

$$\frac{d\hat{O}}{d\boldsymbol{\theta}} = -\frac{\partial O}{\partial \boldsymbol{U}} \left(\frac{\partial \boldsymbol{C}}{\partial \boldsymbol{U}}\right)^{-1} \frac{\partial \boldsymbol{C}}{\partial \boldsymbol{\theta}} + \frac{\partial O}{\partial \boldsymbol{\theta}}, \tag{II.4}$$

where the first term can be evaluated either by solving the adjoint PDE first (from left to right) or solving the tangent linear PDE first (from right to left). For the adjoint method, the adjoint PDE is formulated as follows

$$\frac{\partial C^*}{\partial U}\lambda = \frac{\partial O^*}{\partial U}, \tag{II.5}$$

where $\lambda \in \mathbb{R}^N$ represents the adjoint variable. This is a linear PDE to solve, which requires the solution vector $U$ to get the Jacobian matrix $\frac{\partial R}{\partial U}$. Then we can obtain total derivative as follows

$$\frac{d\hat{O}}{d\theta} = -\lambda^*\frac{\partial C}{\partial \theta} + \frac{\partial O}{\partial \theta}. \tag{II.6}$$

Compared to AD to compute these derivatives, the above expressions of derivatives are hard to be derived by hands, especially when consider the non-linearity of PF.

**Supplementary Note 3: Benchmark Tests**

The performance comparisons of benchmarks between `JAX-PF` and `PRISMS-PF` are based on the code released on PRISMS's website, which are the fundamental models used for solidification and solid-state phase transformation. For the Allen-Cahn equation, $f_{chem}(\eta) = 4\eta \times (\eta - 1) \times (\eta - 0.5)$. The model parameters used in the calculations are $M_\eta = 1$ and $\kappa_\eta = 2$. For the Cahn-Hilliard equation, $f_{chem}(c) = 4c \times (c - 1) \times (c - 0.5)$. The model parameters used in the calculations are $M_c = 1$ and $\kappa_c = 1.5$. For coupled Allen-Cahn and Cahn-Hilliard equations, $f_\alpha(c) = -1.6704 - 4.776c + 5.1622c^2 - 2.7375\ c^3 + 1.3687\ c^4$ , $f_\beta(c) = 5.0c^2 - 5.9746c - 1.5924$, and interpolation function $h(\eta) = 10.0n^3 - 15.0\ n^4 + 6.0n^5$. The model parameters used in the calculations are $M_c = 1$, $M_\eta = 150$, and $\kappa_\eta = 0.3$. For all these three benchmarks, the domain geometry is a cube spanning from 0 to 100 along each axis. No-flux boundary conditions are applied on all boundaries. For both implicit and explicit benchmark testing, they shared the same initial conditions, as shown in Fig. 1. The element size and time step size used in these simulations vary, as described in the section "Results", in order to compare simulations at different resolutions and parallel scaling.

For the benchmark of Eshelby inclusion for lattice misfit in solid-state phase transformations, Young's modulus $E$ equals to 22.5 MPa and Poisson's ratio $v$ equals to 0.3. The domain geometry is still a cube spanning from 0 to 100 along each axis. A second phase is defined at the left-bottom corner, as shown in the Fig. 1. We assume the diagonal elements of $\boldsymbol{\varepsilon}^0$ taken the form:

$$\varepsilon_{ij}^0 = m(\frac{1}{2} + \frac{1}{2}\tanh{(l(r-a))}), \qquad (\text{III.1})$$

where $m$ is the magnitude of the misfit strain inside the inclusion, $l$ determines the thickness of the interface between the inclusion and the matrix, $r$ is the distance from the origin, and $a$ is the radius of the inclusion. The off-diagonal elements of $\varepsilon_{ij}^0$ are zero.

**Supplementary Note 4: Platform Report**

Here, we report the platforms for those numerical experiments. `JAX-PF` (GPU) runs on Intel(R) Xeon(R) Gold 6230 CPU @ 2.10GHz (Cascade Lake) with NVIDIA A100 GPU (40 GB Graphics memory). `PRISMS-PF` CPU with MPI runs on 2.2 GHz AMD Ryzen Threadripper PRO 3975WX (32 Cores).

## Supplementary Note 5: Crystal Plasticity Finite Element Method

The kinematics of isothermal finite deformation describes the process where a body originally in a reference configuration, $B \subset R^3$, is deformed to the current configuration, $S \subset R^3$, by a combination of externally applied forces and displacements over a period of time [111]. In this treatment, we choose the perfect single crystal as the reference state, which has the advantage of a constantly unchanged reference state. The balance of momentum in reference configuration (ignoring inertial term and body force) is expressed as follows:

$$\text{Div } \boldsymbol{P} = 0 \quad \text{in } \Omega,$$
$$\boldsymbol{u} = \boldsymbol{u}_D \quad \text{on } \Gamma_D, \qquad \text{(V.1)}$$
$$\boldsymbol{P} \cdot \boldsymbol{n} = \boldsymbol{t} \quad \text{on } \Gamma_N,$$

where $\boldsymbol{P}$ is the first Piola-Kirchhoff stress tensor, $\boldsymbol{u}$ is the displacement field to be solved, $\boldsymbol{u}_D$ is the boundary displacement, $\boldsymbol{t}$ is the traction and $\boldsymbol{n}$ is the outward normal vector.

The deformation gradient $\boldsymbol{F}$ is assumed to be multiplicatively decomposed in its elastic and plastic parts [112]:

$$\boldsymbol{F} = \boldsymbol{F}^e \boldsymbol{F}^p, \qquad \text{(V.2)}$$

where $\boldsymbol{F}^e$ is the elastic deformation gradient induced the reversible response of the lattice to external loads and displacements, and $\boldsymbol{F}^p$ is the plastic deformation gradient, an irreversible permanent deformation that persists when all external forces and displacements that produce the deformation are removed.

The total plastic velocity gradient can be expressed in terms of the plasticity deformation gradient as

$$\boldsymbol{L}^p = \dot{\boldsymbol{F}}^p (\boldsymbol{F}^p)^{-1}. \qquad \text{(V.3)}$$

The elastic Lagrangian strain $\boldsymbol{E}^e$ is defined as

$$\boldsymbol{E}^e = \frac{1}{2}(\boldsymbol{F}^{eT} \boldsymbol{F}^e - \boldsymbol{I}). \qquad \text{(V.4)}$$

The second Piola-Kirchhoff stress $\boldsymbol{S}$ is given by

$$\boldsymbol{S} = \mathbb{C} : \boldsymbol{E}^e, \qquad \text{(V.5)}$$

where $\mathbb{C}$ is the elastic modulus. The Cauchy stress $\boldsymbol{\sigma}$ is given by

$$\boldsymbol{\sigma} = \frac{1}{\det(\boldsymbol{F}^e)} \boldsymbol{F}^e \, \boldsymbol{S} \, \boldsymbol{F}^{eT}. \qquad \text{(V.6)}$$

The first Piola-Kirchhoff stress $\boldsymbol{P}$ is given by

$$\boldsymbol{P} = \det(\boldsymbol{F})\,\boldsymbol{\sigma}\,\boldsymbol{F}^{-\mathrm{T}}. \qquad (\mathrm{V}.7)$$

The plastic velocity gradient $\boldsymbol{L}^{\mathrm{p}}$ is computed as the sum of contributions from all slip systems.

$$\boldsymbol{L}^{\mathrm{p}} = \sum_\alpha \dot{\gamma}^\alpha\, \boldsymbol{s}_0^\alpha \otimes \boldsymbol{m}_0^\alpha, \qquad (\mathrm{V}.8)$$

where $\dot{\gamma}^\alpha$ is the slip rate for slip system $\alpha$, $\boldsymbol{s}_0^\alpha$ and $\boldsymbol{m}_0^\alpha$ are unit vectors describing the slip direction and the normal to the slip plane of the slip system in the reference configuration. The resolved shear stress $\tau^\alpha$ is defined as

$$\tau^\alpha = \boldsymbol{S} : \boldsymbol{s}_0^\alpha \otimes \boldsymbol{m}_0^\alpha. \qquad (\mathrm{V}.9)$$

The slip rate $\dot{\gamma}^\alpha$ is expressed as a power law relationship:

$$\dot{\gamma}^\alpha = \dot{\gamma}_0 \left|\frac{\tau^\alpha}{g^\alpha}\right|^{1/m} \mathrm{sign}(\tau^\alpha), \qquad (\mathrm{V}.10)$$

where $\dot{\gamma}_0$ is a reference slip rate, $g^\alpha$ is the slip resistance (or critical resolved shear stress), and $m$ is the strain rate sensitivity exponent. The rate of slip resistance $\dot{g}^\alpha$ is given by

$$\dot{g}^\alpha = \sum_\beta h^{\alpha\beta}\, |\dot{\gamma}^\beta|, \qquad (\mathrm{V}.11)$$

where $h^{\alpha\beta}$ is the rate of strain hardening on slip system $\alpha$ due to a shearing on the slip system $\beta$. Kalidindi et al. [88] self and latent hardening law gives

$$h^{\alpha\beta} = q^{\alpha\beta} h_0 \left|1 - \frac{g^\alpha}{g_{\mathrm{sat}}}\right|^a \mathrm{sign}\left(1 - \frac{g^\beta}{g_{\mathrm{sat}}}\right). \qquad (\mathrm{V}.12)$$

Respectively, $g_{\mathrm{ini}}$ is the initial slip resistance and $g_{\mathrm{sat}}$ is the saturation slip resistance. $a$ and $h_0$ are slip hardening parameters which are taken to be identical for all slip systems. $q^{\alpha\beta}$ is the matrix describing the latent hardening behavior of a crystallite. Here,

$$q^{\alpha\beta} = \begin{cases} 1 & \text{if } \alpha = \beta \\ r & \text{if } \alpha \neq \beta \end{cases}, \qquad (\mathrm{V}.13)$$

where $r$ is the ratio of the latent hardening rate to the self-hardening rate.

Materials parameters were sourced from the `MOOSE` benchmark and literature [88].

Table. V. 1 Material Parameters used for three cases

|  | Unit | **Copper** [88] |
|---|---|---|
| Crystal Structure | - | FCC |
| Initial flow stress $g_{ini}$ | MPa | 60.8 |
| Coefficient for hardening $a$ | - | 2.5 |
| Hardening constants $h_0$ | MPa | 541.5 |
| Saturated slip strength $g_{sat}$ | MPa | 109.8 |
| Rate sensitivity exponent $n$ | - | 0.1 |
| Latent hardening coefficient $r$ | - | 1.0 |
| Reference strain rate $\dot{\gamma}_0$ | s$^{-1}$ | 0.001 |
| Elasticity modulus $C_{11}$ | MPa | 168,400 |
| Elasticity modulus $C_{12}$ | MPa | 121,400 |
| Elasticity modulus $C_{44}$ | MPa | 75,400 |

**Supplementary Note 6: Finite-difference numerical derivatives**

Phase-Field Method entails complex nonlinear constitutive relations, and obtaining the sensitivity through analytical derivations using the chain rule is highly challenging, labor-intensive, and prone to errors. A possible approach is based on numerical differentiation such as finite-difference-based numerical derivatives (FDM), which is a kind of numerical method for approximating the value of derivatives. Assuming a small perturbation of the model parameters, the model can be linearized with respect to the perturbation and the new material response as follows

$$\hat{O}(\boldsymbol{\theta} + \boldsymbol{\delta}) \approx \hat{O}(\boldsymbol{\theta}) + \boldsymbol{J}\boldsymbol{\delta}, \tag{VI.1}$$

where

$$\boldsymbol{J} = \frac{\partial \hat{O}(\boldsymbol{\theta})}{\partial \boldsymbol{\theta}}, \tag{VI.2}$$

$\boldsymbol{J}$ is the gradient, namely, the derivatives of material response with respect to the set of parameter variables. To compute the Jacobian matrix, one of the parameters is perturbed by $\Delta\theta$ as follows

$$\boldsymbol{\theta}_j^* = \boldsymbol{\theta} + [0, 0, \dots, \Delta\theta_j, \dots, 0]^\text{T}, j = 1,2, \dots M \tag{VI.3}$$

and the response of the perturbed model is determined by the FEM analysis of the RVE. This procedure is repeated for each parameter in the model and resulting result is given by,

$$J_j = \frac{\partial \hat{O}(\boldsymbol{\theta})}{\partial \theta_j} \approx \frac{\hat{O}(\boldsymbol{\theta}^{*j}) - \hat{O}(\boldsymbol{\theta})}{\Delta\theta_j}. \tag{VI.4}$$

The FDM is easy to implement. However, it is computationally expensive because it needs to execute the forward PF simulation multiple times, especially when the size of the input parameters is larger than that of the objective function (e.g., $M \gg 1$ in this scenario). In addition, when accurate sensitivities are critical, such as calibration of material parameters for quantitative PF models, numerical approximations can lead to substantial errors, which may result in inaccurate conclusions [44].